\documentclass[pra,twocolumn,showpacs]{revtex4}
\usepackage{epsfig}
\usepackage{amsbsy}
\usepackage{amsmath}
\usepackage{latexsym}

\begin{document}

\title{Effects of staggered magnetic field on entanglement in the anisotropic $XY$ model}
\author{Zhe Sun, XiaoGuang Wang, and You-Quan Li}
\affiliation{Zhejiang Institute of Modern Physics, Department of
Physics, Zhejiang University, HangZhou 310027, China}

\date{\today}
\begin{abstract}
We investigate effects of staggered magnetic field on thermal
entanglement in the anisotropic $XY$ model. The analytic results
of entanglement for the two-site cases are obtained. For the
general case of even sites, we show that when the anisotropic
parameter is zero, the entanglement in the $XY$ model with a
staggered magnetic field is the same as that with a uniform
magnetic field.
\end{abstract}
 \maketitle
\section{Introduction}
Recently, the study of entanglement have received more and more
attention, not only because it is one of the most intriguing
properties of quantum physics but also because it plays an
important role in the quantum information
processing~\cite{Nielsen}. An important emerging field is the
quantum entanglement in condensed matter systems such as spin
chains~\cite{M_Nielsen}-\cite{M_Toth}, and it is believed that
entanglement is a signature of critical point in quantum phase
transitions~\cite{QPT_Nature}-\cite{QPT_GVidal}.

Experimentally, entangled state of magnetic dipoles has been found
be to crucial to describing magnetic behaviours in a quantum spin
system~\cite{Exp}. The study of the entanglement structure in spin
chains will be of importance as the entanglement underlies
operations of quantum computing and quantum information
processing. Moreover, the spin chains not only display rich
entanglement features, but also have useful applications such as
the quantum state transfer~\cite{M_Sub}.

It was found that entanglement can be increased by applying an
external magnetic field~\cite{M_Arnesen}, and it was further shown
that in a two-qubit Heisenberg $XX$ model, the entanglement could
be enhanced under a nonuniform magnetic field~\cite{M_Sun}. There
are a lot of studies on nonuniform staggered magnetic field
effects in condensed matter physics. Here, we consider the effects
of staggered magnetic fields on entanglement in the multi-qubit
anisotropic $XY$ model.

We first give analytical expressions of entanglement for the
two-qubit case in Sec.~II, and also give some numerical results.
Then in Sec.~III, we give numerical results of entanglement up to
ten sites, and prove that when the anisotropic parameter is zero,
the entanglement in the $XY$ model with a staggered magnetic field
is the same as that with a uniform magnetic field. We conclude in
Sec.~IV.

\section{Two-qubit model}
We first introduce the concept of negativity, which will be used
as the entanglement measure. The Peres-Horodecki
criterion~\cite{PH} gives a qualitative way for judging if the
state is entangled or not. The quantitative version of the
criterion was developed by Vidal and Werner~\cite{Vidal}. They
presented a measure of entanglement called negativity that can be
computed efficiently, and the negativity does not increase under
local manipulations of the system. The negativity of a state
$\rho$ is defined as
\begin{equation}
{\cal N(\rho)}=\sum_i|\mu_i|,
\end{equation}
where $\mu_i$ is the negative eigenvalue of $\rho^{T_1}$, and $T_1$ denotes \\
the partial transpose with respect to the first system. The
negativity ${\cal N}$ is related to the trace norm of $\rho^{T_1}$
via~\cite{Vidal}
\begin{equation}
{\cal N(\rho)}=\frac{\|\rho^{T_1}\|_1-1}{2},
\end{equation}
where the trace norm of $\rho^{T_1}$ is equal to the sum of the
absolute values of the eigenvalues of $\rho^{T_1}$.

The Hamiltonian $H_1$ for the anistropic $XY$ model with a
staggered magnetic field is given by the following form:
\begin{align}
H_1(\gamma,B)=&\sum_{i=1}^NJ\big[s_{ix} s_{i+1,x}+\gamma s_{iy}
s_{i+1,y}+(-1)^{i-1}Bs_{iz}\big],\nonumber\\
H_2(\gamma,B)=&\sum_{i=1}^NJ\big[s_{ix} s_{i+1,x}+\gamma s_{iy}
s_{i+1,y}+Bs_{iz}\big],\label{h}
\end{align}
where ${\bf s}_i$ is the spin 1/2 operator on the $i$-th site,
$\gamma$ is the anisotropic parameter, $B$ is the magnitude of the
applied magnetic field on the $i$-th spin, and $J$ is the exchange
constant, which is assume to be one (antiferromagnetic case). For
comparison, we also give the Hamiltonian $H_2$ with a uniform
magnetic field. We have assumed periodic boundary condition in the
above Hamiltonians.

We study entanglement of states of the system at thermal
equilibrium described by the density operator $\rho(T)=\exp(-\beta
H)/Z$, where $\beta=1/k_BT$, $k_B$ is the Boltzmann's constant
($k_B$ is set to 1 in the following), and
$Z=\text{Tr}\{\exp(-\beta H)\}$ is the partition function. The
entanglement in this thermal state is called thermal entanglement.
Due to the $Z_2$ symmetry, i.e.,
\begin{equation}
[H,\sigma_{1z}\otimes\sigma_{2z}\otimes\cdots\otimes\sigma_{Nz}]=0,
\end{equation}
the two-qubit reduced density matrix for qubits $i$ and $i+1$ is
given by
\begin{equation}
\rho_{i,i+1}=\frac{1}{Z}\left(
\begin{array}{llll}
a_1&0  &0&  b_1\\
0&  a_2&b_2&0\\
0&  b_2&a_3&0\\
b_1&0  &0  &a_4
\end{array}\right)\label{rho1}
\end{equation}
in the standard basis  $\{|0 0\rangle,|0 1\rangle,|1 0\rangle,|1
1\rangle\}$. After the partial transpose, equivalent to exchanging
$b_1$ and $b_2$ in the above equation, we obtain the expression of
negativity as
\begin{align}
{\cal N}_{i,i+1}=&\frac{1}2\max \big[0,\sqrt{(a_1-a_4)^2+4b_2^2}-a_1-a_4)\big]\nonumber\\
+&\frac{1}2\max\big[0,\sqrt{(a_2-a_3)^2+4b_1^2}-a_2-a_3\big],\label{N}
\end{align}
which is the general expression of negativity for two qubits in
the multi-qubit model.

Now we consider the two-qubit case to obtain some analytical
expression for entanglement. Explicitly, the Hamiltonians with a
uniform and staggered magnetic fields are given by
\begin{eqnarray}
H_1&=&s_{1x}\otimes s_{2x}+\gamma s_{1y}\otimes s_{2y}+Bs_{1z}-Bs_{2z},\nonumber \\
H_2&=&s_{1x}\otimes s_{2x}+\gamma s_{1y}\otimes
s_{2y}+Bs_{1z}+Bs_{2z},\label{h11111}
\end{eqnarray}
respectively. There exists the $Z_2$ symmetry in the above
Hamiltonians, and in the standard basis, and they can be written
in the block-diagonal form. Therefore, the density operator
describing the thermal state $\rho_T$ is also in the
block-diagonal form.

After exponential expansion, for Hamitonian $H_1$, we have the
density operator $\rho_T$ given by the form of Eq.~(\ref{rho1})
with matrix elements
\begin{eqnarray}
a_1&=&a_4=\cosh[\beta(1-\gamma)/4],\nonumber\\
a_2&=&\cosh(\beta D_2/4)+(4B/D_2)\sinh(\beta D_2/4),\nonumber\\
a_3&=&\cosh(\beta D_2/4)-(4B/D_2)\sinh(\beta D_2/4),\nonumber\\
b_1&=&-\sinh[\beta(1-\gamma)/4],\nonumber\\
b_2&=&-[(1+\gamma)/D_2]\sinh(\beta D_2/4),\nonumber\\
D_2&=&\sqrt{16B^2+(1+\gamma)^2}. \label{e2}
\end{eqnarray}
For Hamiltonian $H_2$, the density operator $\rho_T$ is still
given by the form of Eq.~(\ref{rho1}) with the following matrix
elements
\begin{eqnarray}
a_1&=&\cosh(\beta D_1/4)+(4B/D_1)\sinh(\beta D_1/4),\nonumber\\
a_2&=&a_3=\cosh[\beta(1+\gamma)/4],\nonumber\\
a_4&=&\cosh(\beta D_1/4)-(4B/D_1)\sinh(\beta D_1/4),\nonumber\\
b_1&=&-[(1-\gamma)/D_1]\sinh(\beta D_1/4),\nonumber\\
b_2&=&-\sinh[\beta(1+\gamma)/4],\nonumber\\
D_1&=&\sqrt{16B^2+(1-\gamma)^2} .\label{e1}
\end{eqnarray}
Substituting the above equations into Eq.~(\ref{N}), we then get
analytical expressions of the negativity.

\begin{figure}
\includegraphics[width=0.45\textwidth]{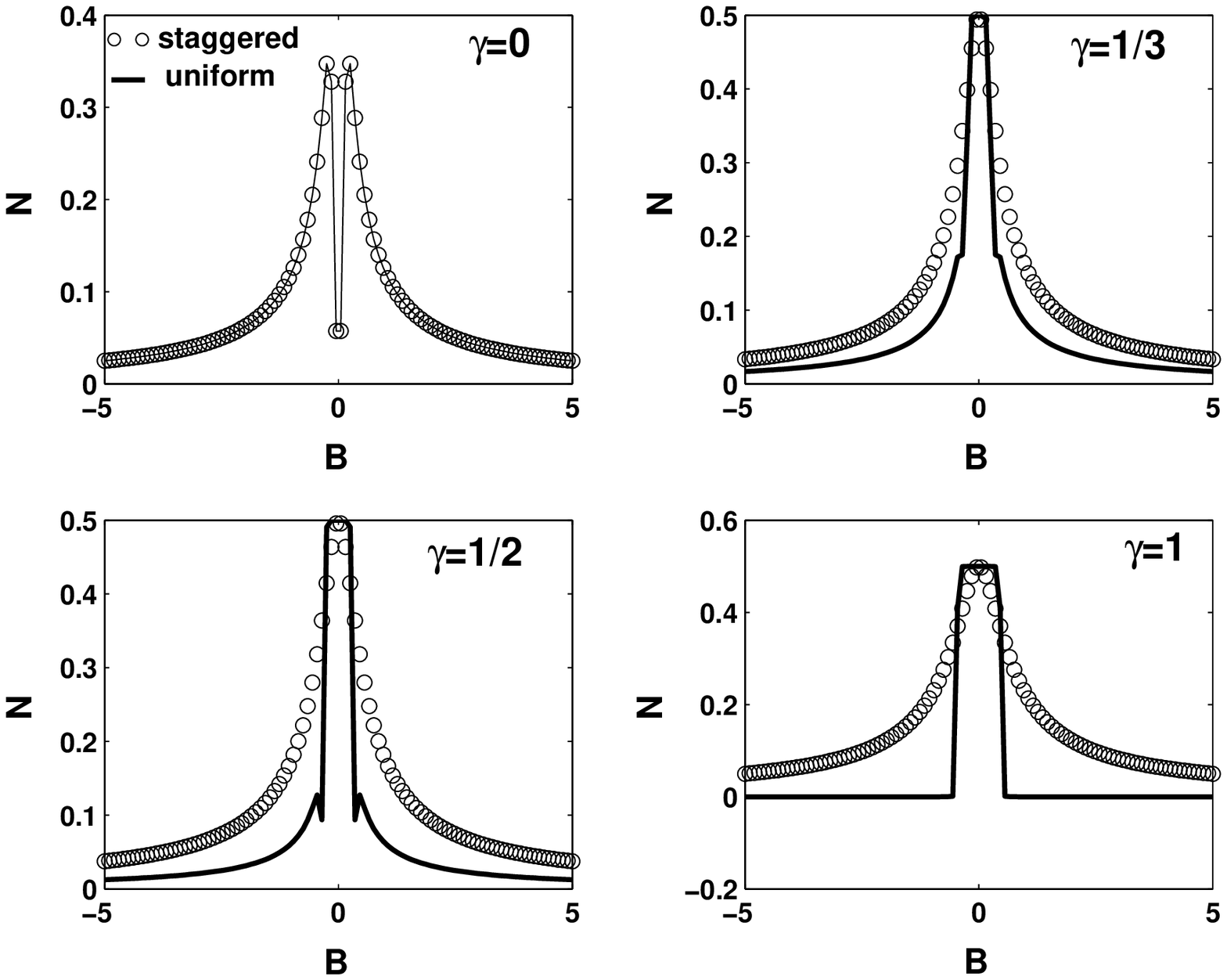}
\caption{The negativity versus the magnetic fields for the case of
two sites at a temperature of $T=0.02$.}
\end{figure}
\begin{figure}
\includegraphics[width=0.45\textwidth]{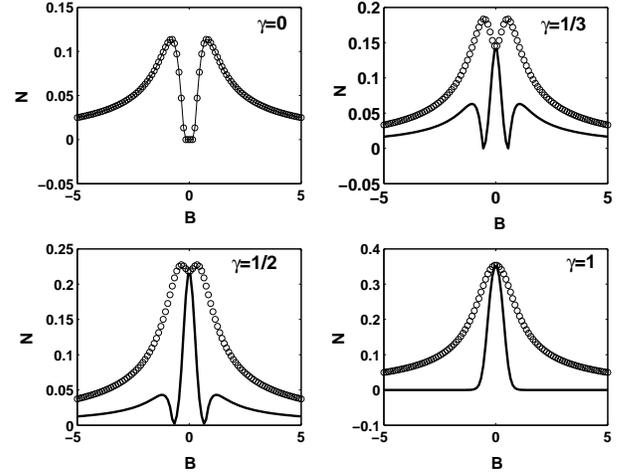}
\caption{The negativity versus magnetic fields for the case of two
sites at a temperature of $T=0.2$. }
\end{figure}

From the analytical results of the negativity, we plot the
entanglement versus $B$ at a lower temperature in Fig.~1 for both
the cases of the uniform and staggered fields. Throughout the
paper, we use circle lines to plot the negativity versus $B$ for
the case of staggered field, and solid lines for the case of
uniform field. We see that the negativity is symmetric with
respect to the magnetic field $B=0$, namely, $\pm B$ gives the
same value of negativity. From the analytical result, changing $B$
to $-B$ results in exchanging $a_2$ and $a_3$ in Eq.~(\ref{e2})
and exchanging $a_1$ and $a_4$ in Eq.~(\ref{e1}). Then, from
Eq.~(\ref{N}), we know that the negativity is invariant when
changing $B$ to $-B$.

For finite anisotropic parameters, the negativity behaves
differently under the two kinds of external magnetic fields. It is
interesting to see that the curve for the uniform field coincides
with that for the staggered field when $\gamma=0$. It can also be
explained from the analytical result of the negativity. When we
take $\gamma=0$ in the expressions (\ref{e2}) and (\ref{e1}), from
Eq.~(\ref{N}), it is direct to check  that the negativities are
equal for the two cases of uniform and staggered magnetic fields.

Now we consider a higher temperature $T=0.2$, and the numerical
results are given in Fig.~2.  When $\gamma=0$, we still observe
that the curve for the uniform field coincides with that for the
staggered magnetic field. This is a general feature for our
system, as will be shown in the following section. For $\gamma\neq
0$, the uniform field and staggered field displays different
effects on entanglement, and the staggered field enhances
entanglement in comparison with the case of uniform field. For
$\gamma=1/3$ and $\gamma=1/2$, the negativity for the staggered
field shows double peaks, while it shows triple peaks for the
uniform field. This feature is dependent on the anisotropy. When
$\gamma=1$, there exists only one peak as was already shown in
Ref.~\cite{M_Sun}. How the staggered field affects entanglement in
comparison with the case of uniform field relies on the
anisotropic parameter. Next, we go beyond the two-qubit case, and
displays some general features of entanglement in the multi-qubit
anisotropic model.

\section{Multi-qubit model}

To consider the multi-qubit model, for convenience, we restrict
ourselves to the case of even number of sites.  In Fig. 3, We
numerically plot the negativity versus $B$ with $\gamma=0$ for
even number of sites up to ten. Both the uniform and staggered
field effects are considered. In this situation, the models reduce
to the transverse Ising models, and again, we observe that the
entanglement is symmetric with respect to $B$, and these two
fields have same effects on entanglement. For finite anisotropic
parameter $\gamma=0.5$ as show in Fig. 4, the behaviors of the
negativity are qualitatively the same for different number of
sites, namely, the uniform magnetic field leads to triple-peak
structure, while the staggered field leads to the double-peak
structure.

\begin{figure}
\includegraphics[width=0.45\textwidth]{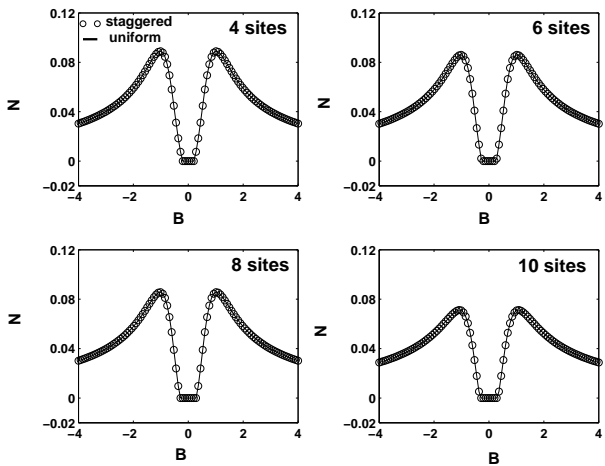}
\caption{The negativity versus the magnetic fields for different
sites. The parameters $\gamma=0$ and $T=0.2$ . }
\end{figure}

\begin{figure}
\includegraphics[width=0.45\textwidth]{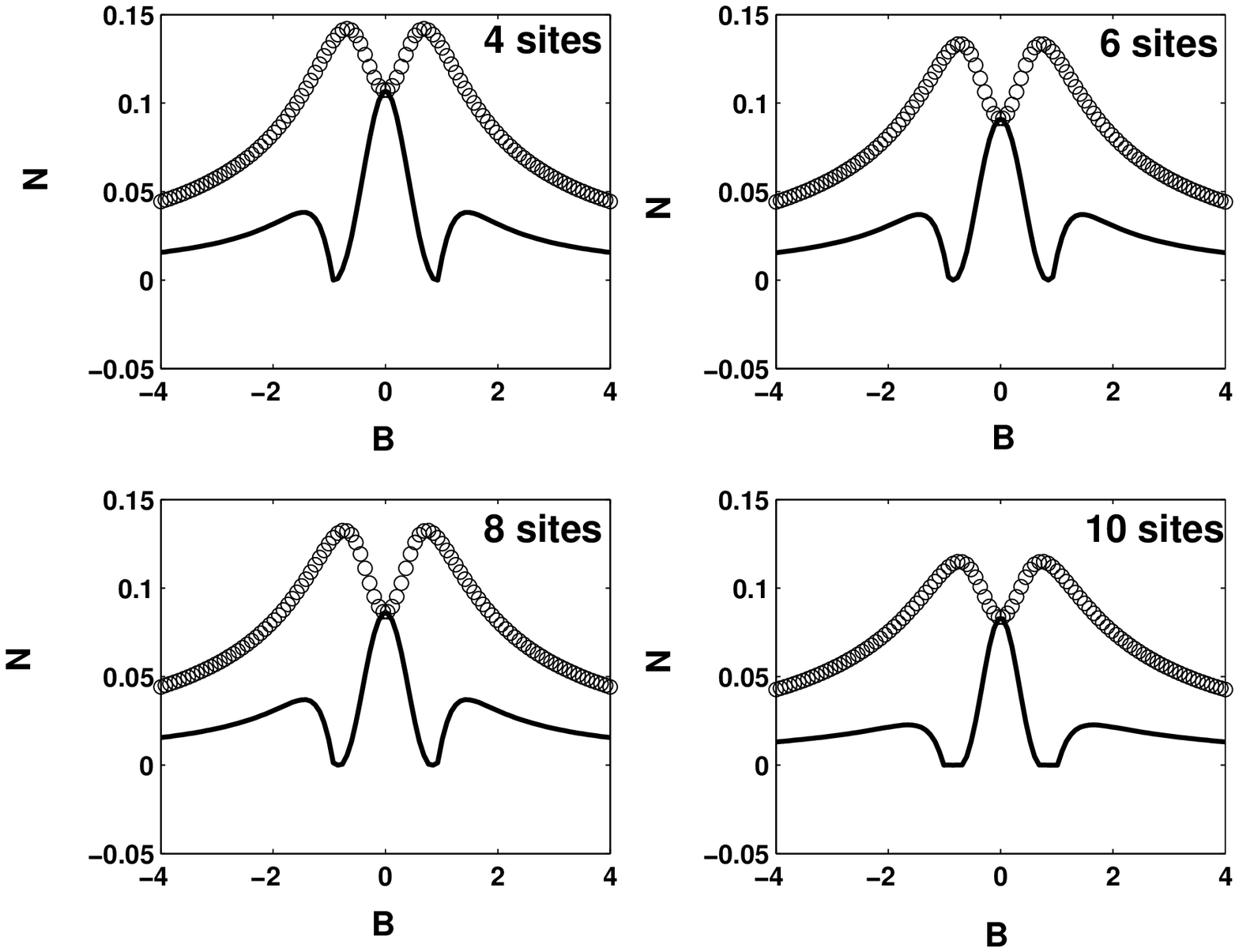}
\caption{The negativity versus the magnetic fields for different
sites. The parameters $\gamma=0.5$ and $T=0.2$ . }
\end{figure}

We first give a strict way to show that the entanglement is
symmetric with respect to $B=0$. Consider the following unitary
transformation
\begin{equation}
U_1=\sigma_{1x}\otimes\sigma_{2x}\otimes\sigma_{3x}\otimes...\otimes\sigma_{Nx}.
\end{equation}
Then, we have
\begin{align}
U_1 H_1(\gamma,B) U_1=&H_1(\gamma,-B),\nonumber\\
U_1 H_2(\gamma,B) U_1=&H_2 (\gamma,-B).
\end{align}
From the above equation, we obtain
\begin{align}
U_1 e^{-\beta H_1(\gamma,B)} U_1=&e^{-\beta H_1(\gamma,-B)},\nonumber\\
U_1 e^{-\beta H_2(\gamma,B)} U_1=&e^{-\beta H_2(\gamma,-B)}.
\end{align}
The thermal state is described by the density operator
$\rho(T)=\exp(-\beta H)/Z$. The partition function is invariant
under a unitary transformation of the Hamiltonian. So, the thermal
state with parameter $B$ is connected with that with parameter
$-B$ via the unitary transformation $U_1$. Another important fact
is that $U_1$ is a {\em local} unitary operator, which will not
change entanglement. Thus, the entanglement is symmetric with
respect to $B=0$.

Second, we show that the uniform and staggered magnetic fields
have same effects on entanglement when the anisotropic parameter
$\gamma=0$. Consider the following transformation
\begin{equation}
U_2=\sigma_{1x}\otimes\sigma_{3x}\otimes\sigma_{5x}\otimes...\otimes\sigma_{N-1,x}.
\end{equation}
Then, we have
\begin{equation}
U_2 H_1(\gamma,B) U_2=H_2(-\gamma,B),
\end{equation}
indicating that $U_2$ transform the Hamiltonian $H_1$ to $H_2$
with $\gamma$ being changed to $-\gamma$. This unitary operator is
a local unitary operator, and thus will not change the
entanglement. Specifically, when $\gamma=0$, we have the equality
$U_2 H_1(0,B) U_2=H_2(0,B)$, and thus the two magnetic fields have
the same effects on entanglement.
\section{Conclusion}
In conclusion, we have studied the entanglement properties of the
anisotropic spin-half system in a staggered external magnetic
field, and compared it with the case of a uniform magnetic field.
We have investigated the generic $XY$ Heisenberg model with
different anisotropy $\gamma$, and  have obtained the analytic
results of negativity in two sites for arbitrary $\gamma$, which
helped us to explain  entanglement properties. We strictly show
that for any temperature the entanglement is symmetric with
respect to zero magnetic field, and when $\gamma=0$, the
negativity for the case of staggered field is the same as that for
the uniform field. If $\gamma\neq 0$ the two kinds of magnetic
fields have different effects on entanglement. We have found that
the staggered magnetic field leads to higher entanglement and
double-peak structure. Here, we have considered the anisotropic
$XY$ model, it will be interesting to consider staggered magnetic
effects on entanglement in other physical magnetic models.

\end{document}